\documentclass[sigconf]{acmart}

\usepackage[english]{babel}
\usepackage{blindtext}
\usepackage{changepage}
\usepackage{hhline}
\usepackage{subfigure}
\usepackage{longtable}
\usepackage{algorithm}
\usepackage{algpseudocode}
\usepackage[section]{placeins}
\usepackage{enumitem}

\def\startfigure{\vspace{6pt}\begin{figure}[ht]\center}

\renewcommand\footnotetextcopyrightpermission[1]{} 
\setcopyright{none}

\acmDOI{}

\acmISBN{}

\acmPrice{}

\begin{document}
\title{LEO Satellite Networking Relaunched: Survey and Current Research Challenges}


\author{Cedric Westphal, Lin Han, Richard Li}
\email{{cedric.westphal,lhan, rli}@futurewei.com}

\renewcommand{\shortauthors}{C. Westphal, et al.}

\begin{abstract}
    This document surveys recent and current developments in LEO satellite networking. It presents a brief overview of satellite networking in order to contextualize the issue. It then focuses on current research work in emerging domains, such as Machine Learning, SDN, low latency networking, green networking, Information-Centric Networks, etc. For each, it presents recent works and a direction of the research community within that emerging domain. 
    
    The paper also describes the current state of standardization efforts in 3GPP and in IETF for LEO satellite networking. In particular, we present in some detail the direction these standards body are pointing towards for LEO networking with inter-satellites links. Finally, some future challenges and interesting research directions are described and motivated. This is an overview of the current state of the LEO satellite research in both academic and industrial standardization environments which we believe will be helpful to understand the current state of the art.  
\end{abstract}

\maketitle

\section{Introduction}

There are (as of January 1st, 2023) 6,700 satellites orbiting the Earth~\cite{ucsusa}. 6,000 of the satellites are in a Low Earth Orbit (LEO, \cite{LEO}); roughly 600 are geostationary~\cite{GSO}, and the rest is in between~\cite{MEO}. 
The vast majority of these satellites belong to some commercial deployments (3,000 commercial satellites for the United States alone). Interconnecting these satellites has become an important avenue for both research and development. About 50,000 active satellites are currently planned to orbit overhead before 2030.

China has requested orbit and spectrum resource from ITU for 12,000 satellites~\cite{CHN2020-33634, CHN2020-33636}. 3GPP (3rd Generation Partnership Project) plans to integrate LEO satellites with 5G in its proposed Non-terrestrial-network (NTN) Integration~\cite{TR.38.821} for the future Internet.

Satellite networking as an area of research investigation started to pick up in the 1990s. However, in recent years, some new developments have thrown more fuel on the fire of satellite networking research. In particular, some high profile efforts to get connected through the Internet via LEO satellites have demonstrated the business proposition of creating networks in space, and networks in the sky. 

We attempt here to survey the recent turns taken by satellite networking over the last few years. One key aspect is that communications have evolved from the so-called "bent-pipe model" -- where what the satellite receives from the ground is reflected back to the ground -- to a model with inter-satellite communications -- where a data stream is sent up from the ground to a satellite, and then through a series of relaying satellite, before being sent back to the ground destination. Inter-satellite links use free-space optics to connect fast moving object 2,000 miles apart.

Another key aspect is that major networking and routing innovations would naturally expand from the Internet on the ground to a network in space. For instance, the deployment of SDN has been applied in space as well, as we will discuss below. 

Further, the continuous growth of the number of satellites in LEO, as well as the high cost of deploying satellites, would indicate that some level of interoperability is ineluctable; and therefore, standardization organizations are discussing protocols so that satellites from different vendors could talk to each other. StarLink itself added 457 to its constellation in the first six month of 2023.

\begin{figure}[ht]
\centering
\includegraphics[width=3in]{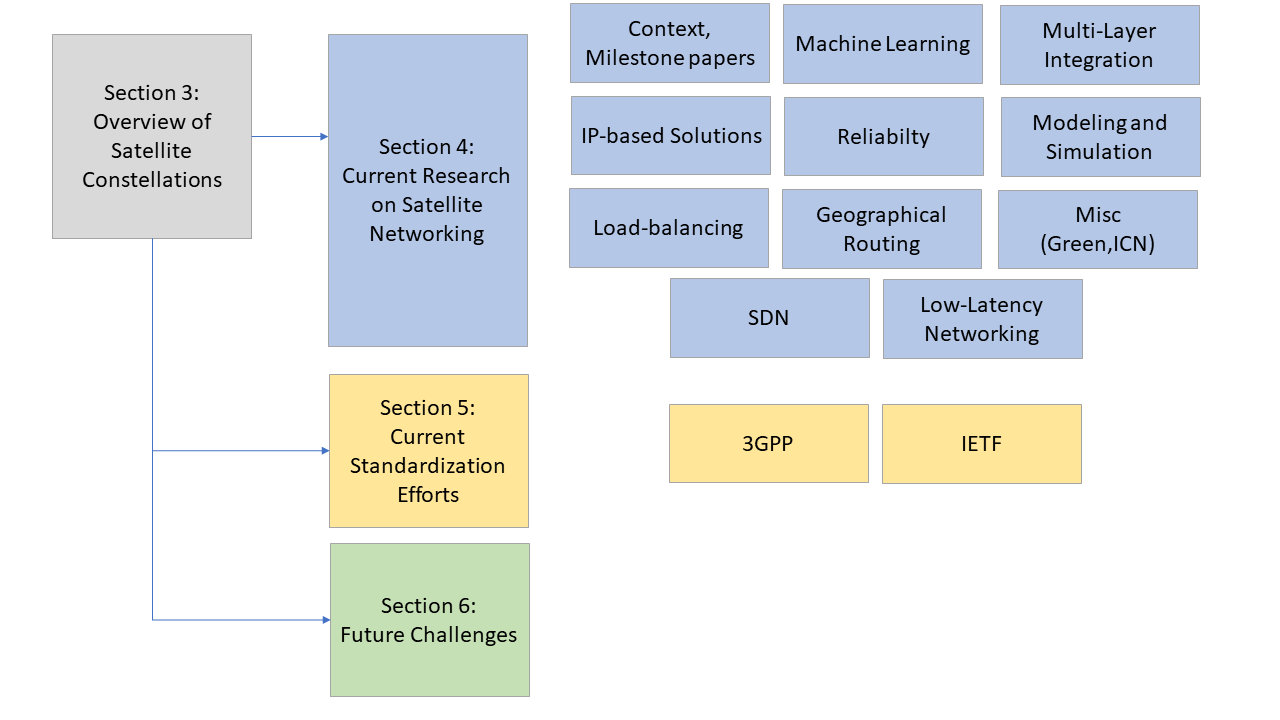}
\caption{Organization of this survey}
\label{fig:organization}
\end{figure}

Our paper is intended as a introductory survey of major recent developments in satellite networking. As such, it is organized as follows: in section~\ref{sec:related}, we start by contrasting this work with the prior research surveys on a similar theme. In Section~\ref{sec:overview}, we frame the problem by discussing Low-Earth Orbit satellites, the constellations, the main commercial deployments. In Section~\ref{sec:routingresearch} we review the main recent research results for routing in LEO satellite networks. This section is composed of different subsections, looking at, among other topics: the milestones in the field; IP-based solutions; the emergence of SDN in satellite networking; the integration of Machine Learning techniques; the focus on low-latency networking; the simulation platforms and models to describe such networks. Section~\ref{sec:standards} considers the effort to standardize protocols for satellite networks, in 3GPP and IETF. Section~\ref{sec:challenges} looks forward to the challenges and research issues to consider in the future; finally Section~\ref{sec:conclusion} concludes the paper. Figure~\ref{fig:organization} presents the organization of the paper. 

\section{Related Works}
\label{sec:related}

As this work surveys the current literature, we only consider previous overviews and surveys as part of the related work. A lot of current research on LEO satellites will be presented in the subsequent sections in an organized manner in Section~\ref{sec:routingresearch}.

Routing in satellites generally maps to two categories: {\em snapshot} routing, and {\em dynamic} routing. With the former, the topology of the satellite constellation is supposed to be static for the duration of a {\em snapshot} and the routing is then performed on this static topology. Due to the different latency for a packet transmission (of the order of milliseconds) and the motion of the satellite (that can be in view of a ground station for several minutes), this assumption is reasonable. As the topology evolves, a new routing table is populated for the next snapshot. Routing tables can typically be pre-computed and t0 hen uploaded (or stored) to minimize the processing workload in the satellite. 

The alternative is dynamic routing, where the satellite forwards packets based upon the current network configuration. This means the routing protocol must dynamically adapt to changes in the topology, as the length of time a snapshot is stable is rather short (a problem similar to characterizing the stability of a path in ad hoc networks~\cite{richard2005defining}). \cite{cao2022dynamic} offers a very recent and very thorough survey of dynamic routing. Our work encompasses a wider approach, focusing on the new trends in satellite networking, and not exclusively on dynamic routing. 

\cite{xiaogang2016survey} surveys satellite networking and routing from the beginning of the field and presents a good overview of the literature starting from the 1990s. It also covers multi-layer satellite routing, combining low-earth orbit with satellite higher up. Our paper is focused on more recent work.

\cite{darwish2021location} looks at location management for satellite networks and offers a comprehensive survey of this topic, including location management in generic networks. They focus on three approaches: the extension of current IETF protocols, the adoption of location/identifier split methods, and the use of SDN for LEO location management. 

IP routing issues in LEO satellite networks were identified early in~\cite{wood2001IP} or~\cite{gounder1999routing}. The issues listed in~\cite{wood2001IP} for instance include: variable IP packet sizes, that may require fragmentation and make the channel allocation more complicated; the complexity of IP route table management in a highly dynamic environment; the overhead of IP routing vs switching in power, computation and time to find a longest prefix match. This has suggested the use of other techniques to alleviate or fix these issues. \cite{taleb2005IP} for instance offers a nice survey of routing in satellite as of 2005, and evaluate the suitability of IP protocols for satellite routing - and in particular, that of TCP. They propose a Recursive, Explicit, and Fair Window Adjustment (REFWA) to tune TCP for use in satellite networks. 

As more and more LEO constellations are being deployed, LEO satellite networking in space has received more attention for research into integrating space and ground networks~\cite{Liu2018SpaceAirGroundIN}.

\cite{alagoz2007exploring} surveys routing strategies in LEO constellations, considering the the particular features of satellite networks, such as dynamic topology, non-homogeneous traffic distribution, limited power and processing capabilities, and high propagation delays. The considered protocols are classified according to their routing objectives, and pros and cons are discussed. 

\cite{guidotti2019architecture} considers the 5G challenges for satellite networks from the point of view of the MAC layer - with some discussion of the physical layer. It is therefore complementary to this paper, which focuses on the networking layer and above. 

\cite{darwish2021LEO} surveys 3GPP and 5G standardization efforts. It is a recent work that overlaps in part with Section~\ref{sec:3GPPPsatellite} but does not discuss other emerging trends in satellite networking. 

In a short survey \cite{gaber20205G} consider opportunities to integrating the terrestrial mobile and satellite networks within 5G. They also propose a mobility management scheme to reduce signaling overhead and to reduce service disruption when performing handoffs in between satellites.

Similarly, \cite{xie2022survey} also briefly surveys satellite mobility management, but from the point of view of 5G and 6G networks. This survey also evaluates 3GPP proposals, some of them are described below in Section~\ref{sec:3GPPPsatellite}.


\section{Satellite Networking: Overview}
\label{sec:overview}

\begin{figure}[ht]
\centering
\includegraphics[width=3in]{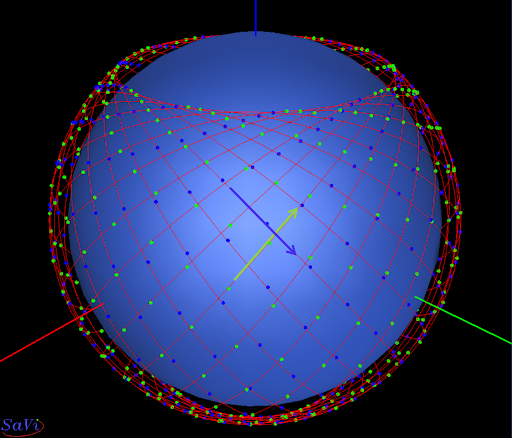}
\caption{Half Satellites (green dots) move on the different direction as the another half Satellites (blue dots) (20 Orbit Planes, 20 Satellites per Orbit Plane)}
\label{fig:sat_move}
\end{figure}

As mentioned before, there are currently 6,700 satellites in Low Earth Orbit, that is, in the altitude range below 2,000km. Orbits at less than 300km of altitude suffer too much drag, so LEO satellites are placed in orbit mostly between 500km and 2,000km.

Geostationary (GEO) satellites are at 36,000km; this is far enough that one satellite can cover a lot more ground. Indeed, four GEO satellites can cover the whole Earth. Also, as their name indicates, their period around the Earth coincides with the 24 hours rotation, and they are viewed as stationary, hovering at the vertical of the same spot on Earth, when observed from a point on the ground. 

\cite{wood2003satellite} offers a good primer on the different types of constellations at different altitudes.

Few geostationary satellites that move very little sounds great. However, the distance from Earth is prohibitive: it takes 120ms for the signal to go from the ground to a GEO satellite, and as much to come back. That is 1/4s just in signal propagation. This is too much for many real time applications. This is what makes LEO satellites attractive for Internet and communications.

It takes only 3ms (at speed of light) to go from the ground to a 1,000km orbit. The shortest path between two points in LEO orbit (at 1,000km orbit) is only 15\% longer than the shortest path between their radial projections on the Earth surface (either when considering the Euclidian distance, or the great-circle/geodesic distance on their respective sphere).

This last point is extremely attractive, considering that the speed of light in fiber is 2/3rd of the speed of light in space; that is, you may add 15\% in distance going over satellite links, but your signal will travel 1.47 times faster than in an optical fiber on the ground, resulting in a net gain\footnote{As a simple rule of thumb, the gain in shorter latency starts from distance on the ground roughly equal to twice the altitude of the satellites. So roughly 1,000km on the ground for LEO at 500km. In practice, the gain may start earlier, as the fibers do not follow the shortest path between the points, but may follow different routes to avoid geographic obstacles, to go under oceans, or avoid unpopulated areas.}. \cite{bhattacherjee2018gearing} provides a detailed assessment of the delay gain. Faster networks are critical to the deployment of new applications (see for instance~\cite{he2018network} for challenges in AR/VR applications).

At LEO altitude, gravity is roughly similar to that on Earth (as the Earth radius is 6,371km) and satellites need speed to stay within that orbit. Then the centrifugal force counteracts the gravity. This means that LEO satellites rotate around the Earth in a period less than 128 minutes. Starlink satellites have a speed relative to the ground of 28,080 km/h. 

Going around the Earth in two hours means that a satellite will not be able to connect to a ground station for an extended period of time. 

Satellites are typically flying at circular orbits, as this allows for constant altitude and therefore constant signal strength to communicate. Figure~\ref{fig:sat_move} displays a typical configuration with 20 orbit planes and 20 satellites per orbit plane.

An {\em orbital shell} is a set of artificial satellites in circular orbits at the same fixed altitude. In the design of satellite constellations, it usually refers to a collection of circular orbits with the same altitude and, oftentimes, orbital inclination, distributed evenly in celestial longitude (and mean anomaly).

\startfigure
\includegraphics[width=\columnwidth]{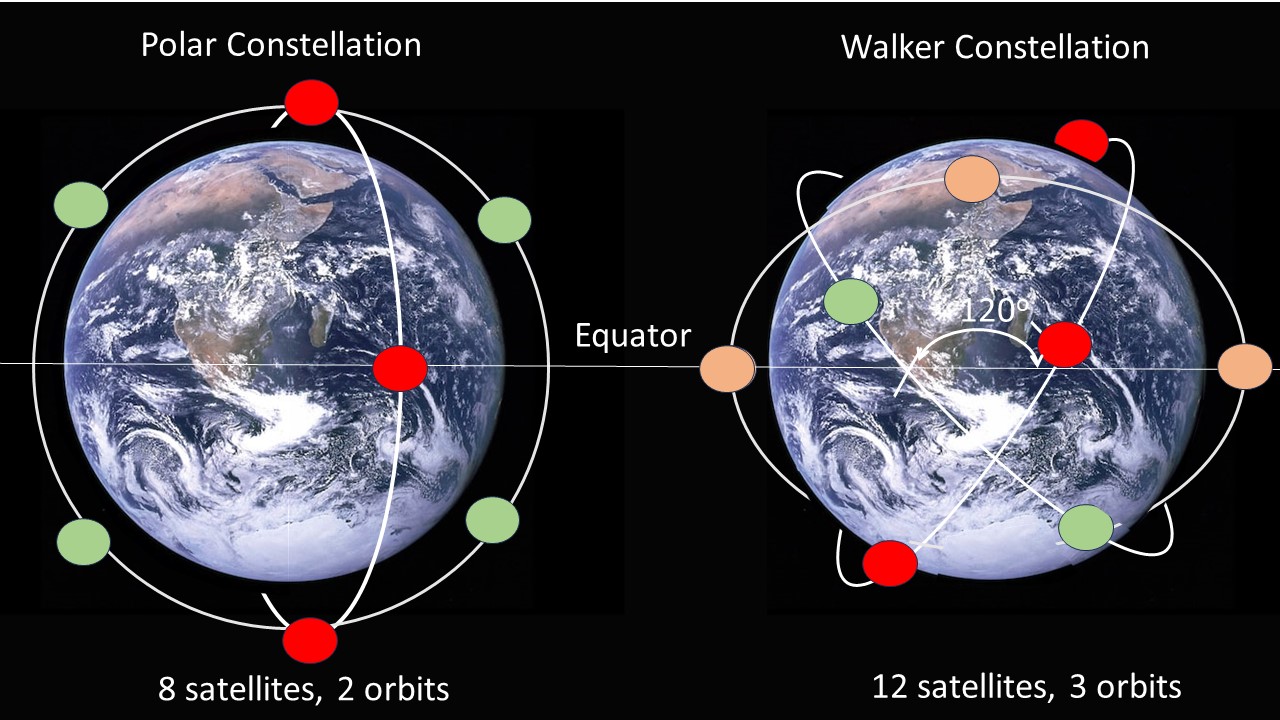}
\caption{Polar Star, Walker Delta (Rosette)}
\label{fig:Walker} 
\end{figure}

Walker~\cite{walker1984constellations} described the Walker constellations (see~Fig.\ref{fig:Walker}. Ballard~\cite{Rosette-satellite} suggested to use a Rosette  constellation that offers invariants for transmissions (also on Fig.\ref{fig:Walker}).

These constellations are parameterized by their altitude (or orbit shell), their inclination and their mean anomaly (see~\cite{Orbit-Elements} for definitions and Fig.~\ref{fig:orbital} for a pictorial description). Raan stands for right ascension of ascending node, namely a parameter that describes the orbital plane trajectory where it intersects the equatorial plane. 

\startfigure
\includegraphics[width=\columnwidth]{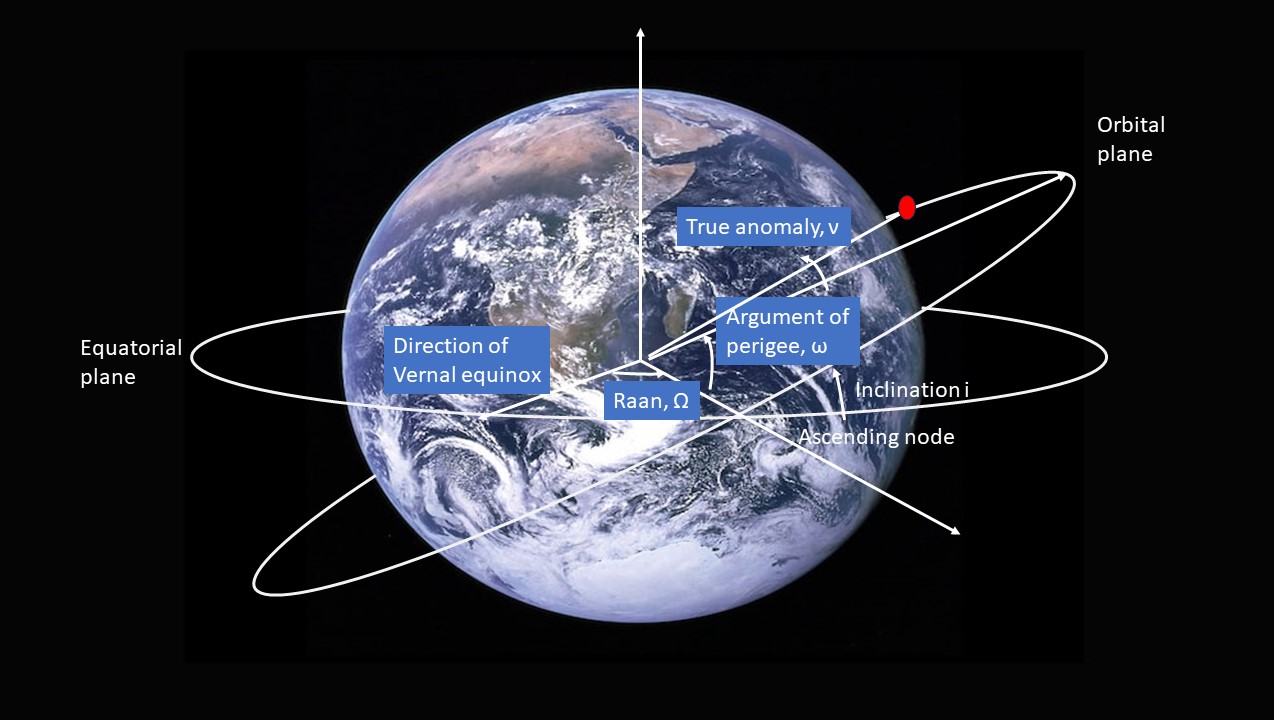}
\caption{Satellite Orbit Parameters}
\label{fig:orbital} 
\end{figure}

\cite{li2021fractal} extends the previous results to construct a F-Rosette constellation that allows inter-satellite links and connectivity~\cite{ISL}. In a similar manner, \cite{Internet-in-Space} takes advantage of the periodic fixed ground trajectory of the satellites (denoted "fixed satellite sub-point trajectory) to specify how to connect satellites with ISLs, define their addresses, and provide routing of traffic among satellites and terrestrial users.

It should be noted that different types of satellite constellations can be combined, by having LEO satellites communicate with MEO or GEO nodes. We focus here only on LEO constellations, but keep in mind that other layers can be sometimes complementary. For instance, a GEO satellite may fill in the coverage blind spots of a LEO constellation, say at the poles. For another example, \cite{chen2002satellite} suggests to use MEO satellites as some sort of controllers for LEO satellites in a hierarchical structure. 

Starlink~\cite{StarLink} is one of the most successful satellite companies right now. Its parent company, SpaceX, is valued at \$150 billion, and its rocket business and innovative satellite engineering enables a relatively cheap deployment for a lot of satellites. As of August 2023, it consists of over 5,000 mass-produced small satellites in low Earth orbit and is the largest constellation in orbit. \cite{handley2018delay} studies the potential performance of a network using the numbers from the Starlink constellation (as of 2018), in terms of the number of orbits, number of satellites per orbit, and the altitude of the satellites. It evaluates the gain in latency in particular between cities such as NYC and London, where a faster-than-fiber connection would provide a business incentive (to the financial industry at least). It then offers a research agenda for the study of such constellations.

\cite{ma2022network} presents a performance study of the actual Starlink network as deployed in 2022. They measured the network performance from different locations in different settings (all within British Columbia). They show that current latency is similar (but not faster) than ground latency; they observe a high rate of outage and a greater sensitivity to weather conditions. 

OneWeb~\cite{OneWeb} and TeleSat~\cite{TeleSat} are providers of broadband Internet service via satellite networks. OneWeb planned to deploy an initial 648-satellite internet constellation in low Earth orbit. At the time of this writing, two third of the satellites have been deployed at an altitude of 1,200kms. TeleSat has currently deployed 188 satellites~\cite{TeleSat2} at an altitude of 1,000km. TeleSat has planned to expand to 1,600 satellites and went through an IPO at the end of 2021 to raise the funding. 

\section{Routing in Satellite Networking: Research Perspective}
\label{sec:routingresearch}

Equipped with the basics of LEO satellite networking, we now turn our attention to the current research on this topic. This section is broken up in eleven subsections, focusing respectively: early progress in LEO satellite networking; IP-based solutions; load balancing; SDN-based solutions; integration of Machine learning; disruption tolerant networking; geographic routing, low-latency networking; multi-layer satellite networking; other research topics; and, simulation platforms and modeling. Each of these topics is motivated within its own section.

\subsection{Early progress in LEO Satellite Networking}
\label{sec:history}

Many of the proposals for satellite networking came in the late 90s, early 2000s, when Iridium, Globalstar, Teledesic hoped to provide communications - and when direct broadcast satellite (DBS) became commercially successful for TV broadcast (with DirecTV and EchoStar). We attempt to capture here the important milestones in this field.

In 1995, \cite{chang1995topological} (later expanded in~\cite{chang1998FSA}) observed that the topology of the satellite constellation followed a limited number of possible states. Therefore, a Finite State Automaton (FSA) could be constructed ahead of time. For each state, a configuration could be pre-calculated and loaded into the satellite that included the solution to a a combined topological design and routing problem. Therefore, the satellite themselves did not have to perform computation, but only to load the topological configuration and routing that corresponds to a specific state. 

\cite{gounder1999routing} defines similar states, denoted as {\em snapshots}, which correspond to the topology of the satellite network at a particular instant of time. Each satellite stores the routing information corresponding to a few snapshots, that are uploaded periodically from the ground stations. The routing table use tag switching (or label switching). This was inspired by ATM proposals at the time. 

ATM itself was modified for satellite networks~\cite{werner1997ATM}. This paper used the ATM Virtual Path Connections (VPCs - logical connections with the same endpoint) to compute paths for a complete period in advance, similar to implementing a set of (time dependent) routing tables.

\cite{jamalipour1997leo} provided a comprehensive overview for using Low Earth Orbit satellite for personal communications already in 1997. 

In 2000, \cite{ekici2000datagram} proposed datagram routing in LEO satellite networks (later expanded into~\cite{ekici2001distributed}). This is a distributed routing protocols that takes into account propagation delay (with the objective of minimizing such delay), congestion and the topology structure. Because it is a per-packet (datagram) forwarding decision, there is no issue of handover when a link breaks, and no connection to maintain. 

This proposal was further expanded to support multicast routing in~\cite{ekici2002multicast}. It seems natural that the features of terrestrial network were progressively adapted to satellite networks. 

\cite{akyildiz2002MLSR} considered multi-layer satellite networks, using LEO, MEO and GEO satellites, as we will describe in Section~\ref{sec:multilayer}.

\cite{papapetrou2007distributed} suggested to use IP and AODV into satellite networks, and proposed a location-assisted on-demand routing (LAOR) protocol. Invoking the shortest path discovery procedure independently for each individual communication request adds a significant overhead, but allows to take into account real time conditions. For a highly dynamic topology, a protocol that allows for some path flexibility around the short path (such as OPRAH~\cite{westphal2006oprah}) may provide benefits over AODV.

\cite{evans2005integration} looked back in 2005 at the history of satellite systems, to argue that while niche services (such as coverage at sea) will continue, satellite systems will have to be integrated with terrestrial networks to be successful. Integration is still an objective today, but there is more confidence today in the deployment of satellite constellation that are competitive - at least in some dimension such as latency - with ground networks. 

\subsection{Focus on IP-based Solutions}
\label{sec:IP}

The Internet is built on a suite of protocols that were designed for ground networks. TCP/IP~\cite{rfc793}, IPv6~\cite{rfc8200} for addressing, BGP~\cite{rfc4271} to share routing information, OSPF~\cite{rfc5340} to compute the routes are commonly use. See~\cite{peterson2011computer} for an overview of Internet protocols. Unfortunately, these are inadequate in the high mobility, highly dynamic environment of satellite constellations. 

For the transport layer, \cite{taleb2005IP}, as discussed in Section~\ref{sec:related} focuses on TCP performance and offers some fixes. 
\cite{makki2003leo} optimizes the routing to stay compatible with TCP; namely they develop a routing algorithm that maximizes the RTT delays compared to the TCP timer granularity. Some innovative transport framework~\cite{li2019framework,li2021qualitative,westphal2021qualitative} may be useful in a constrained satellite network. 

\cite{Internet-in-Space} has analyzed that the network usability will be dramatically reduced to less than 20\% if IP protocols are used as-is. \cite{Internet-in-Space} made proposals to fix this issue, but they only work for F-Rosette satellites~\cite{Rosette-satellite} now and need further work to apply to general LEO satellites.

OSPF is expanded into ASER~\cite{zhang2021ASER}, that is an Area-based SatellitE Routing  protocol. ASER is hierarchical, grouping satellites within an area; within an area, the topology is static (unless unpredictable topology changes happen). ASER is built on top of OSPF to connect these areas dynamically. 

\cite{pan2019OPSPF} also modifies OSPF into an Orbit Prediction Short Path First (OPSPF), taking into account the periodic predictability of the topology. This fixes the issue of endless route convergence with OSPF that consumes expensive inter-satellite link bandwidth.

\cite{luo2021refined} refines Dijkstra's algorithm so as to optimize for satellite networks, by selecting the route with equal cost but fewest route updates (denoted StableRoute). \cite{liu2018suboptimal} also attempts to optimize Dijkstra's shortest path algorithm, but for massive ground and satellite networks without Inter-Satellite Links (ISLs; all other works considered in this paper assume ISLs). 

Other link state routing algorithms include~\cite{zhang2020routing}. The routing algorithm first sets up a topology establishment phase, then a routing calculation phase and finally a link failure response phase. Routing decisions for each hop are based on the link state information with neighboring satellites as well as the network topology.

BGP needs to be modified as well. Back in 2010, \cite{chertov2010using} conducted an experimental study of the link intermittency on the space/ground link and evaluated its impact on BGP peering sessions between ground and satellite routers. The performance degradation is dramatic and \cite{chertov2010using} thus demonstrates the need for a routing architecture that adapts to the particular characteristics of satellite links. 

\cite{giuliari2020internet} asks itself: what would be the performance of existing protocols, such as BGP, in space. It attempts to quantify the impact of such protocols on the performance of the integrated network. It then proposes some clean slate solution to solve the performance gap.

The Border Gateway Protocol – Satellite version (BGP-S)~\cite{ekici2004bgp} is another (earlier) attempt to solve this performance gap by proposing a protocol that allows automated route discovery of the satellite network routes. To simplify the protocol, some simple rules are required (namely that there is only one peer gateway in an AS, and that BGP-S adopts the routing policies constructed by BGP4).

To overcome the limitations of IP in LEO constellations, New IP has been proposed. New IP~\cite{TSAG-C-83,li2018new,li2020new} was introduced to support the evolution of IP as new requirements are introduced by new applications that were unforeseen when IP was designed. For instance, New IP supports flexible addressing, better handle of some QoS parameters so as to support deterministic networking, the support of contracts being carried in the packet, etc. 

One key aspect of New IP to be used in satellite networks~\cite{han2022new} is its ability to use flexible addressing to address satellites based upon their position within an orbit, their orbit and their orbit shell, so as to provide efficient routing. An Internet packet can then be routing over the satellite network using New IP transparently.

\textbf{What next?} We believe there are some tensions in using IP in satellite networks. The protocols designed for the legacy Internet obviously do not anticipate the characteristics of a LEO constellation. Further, current networks are proprietary and do not inter-operate. The key business advantage of, say, StarLink, is that it is able to launch more satellites cheaper than its competitor, and inter-operability would take away their first-mover advantage. On the other hand, inter-operability may be imposed by the regulator, and the integration of the satellite segment into an IP network may become mandatory. 

\subsection{Load balancing}
\label{sec:loadbalancing}

Satellite networks suffer from a very uneven distribution of the traffic. Indeed, to provide global coverage, the satellites are distributed all over the Earth, which has a very uneven density of population. 

A satellite hovering over the Pacific ocean will see little traffic; it may see cross-Pacific traffic, especially if it offers a faster-than-fiber connection between, say, Shanghai and Los Angeles. But very little traffic will originate there. Further, the satellite will follow a trajectory that may take them to areas where they are less critical (say, above the poles, where the distance between satellites shrink when in polar orbit). On the other hand, some links between densely connected areas may see a lot of traffic. 

This, plus the dynamic topology as well as the large number of nodes, creates the need for care in load balancing the traffic. This has been a very active area of investigation. 

\cite{taleb2009explicit} introduced the idea of an explicit exchange of congestion information between neighboring satellites. Nodes inform neighbors when they are about to reach a congestion threshold, and ask them to throttle the traffic towards them. The neighbors then try to route around the congested node. This is called Explicit Load Balancing (ELB). 

\cite{liu2020load} first computes the paths using a modified version of Djikstra's algorithm. This modified algorithm, selective iterative Dijkstra algorithm (SIDA) modifies the traversal of the nodes when computing the shortest path calculation, so as to avoid that multiple shortest paths use the same link (and thereby increase the utilization of that shared link). SIDA tried to avoid repeated use of the same links but keeps the shortest paths of the same length. This is complemented by a selective shunt load balancing (SSLB) strategy. SIDA spreads the traffic over the path, and SSLB reacts to congestion dynamically. 

\cite{liu2017hgl} proposes a hybrid global-local load balancing routing. This takes into account the periodically deterministic nature of the network topology as well, as well as the predictive nature of the considered traffic (here, IoT but this can be generalized to other forms of traffic). This allows to decompose the traffic into a predictable long-term baseline, and some variable short-term fluctuations. The former is managed ahead of time using some global optimization, while the latter is handled dynamically and locally. 

In~\cite{liu2018hybrid}, another hybrid load balancing approach is proposed: they use a prediction of the regional and real-time network states, as input to a multi-path route calculation. The goal is to avoid cascading congestion by balancing traffic onto a path that becomes congested, that again shifts the traffic away. It suggests to shift traffic following a Long-Distance Traffic Detour (LTD) method, so as to avoid overloading neighboring links. This LTD coordinates with distributed traffic detour method to perform self-adaptive load balancing. Both method combine into a Hybrid-Traffic-Detour based Load Balancing Routing (HLBR) scheme.

Another load balancing method~\cite{huang2018load} takes into account the congestion level of the current node as well as the nodes on the path to the destination ground station. The route selection takes into account both congestion level as well as end-to-end delay estimate. This method combines off-line computation and on-line adjustments. (In addition to load balancing, the paper also takes into account - and simplifies- the computation of the route table lookups).

Back-pressure routing is applied to satellite networks in~\cite{deng2022distance}, which proposes a distributed Distance-based Back-Pressure Routing (DBPR). The distance metric used is a combination of shortest path and congestion. The routing is restricted to a rectangle defined by the source and the destination of the traffic, so as to limit hop count and reduce delay. 

\cite{wang2019load} defines a load balancing routing algorithm based on congestion prediction (LBRA-CP). It uses an ant colony algorithm as a heuristics to solve a multi-objective optimization problem, where the objectives include minimizing path costs, but also minimizing congestion (based upon a prediction of the traffic) so as to achieve load balancing. 

\cite{liu2019load} uses segment routing~\cite{rfc8402,rfc8986,ietf-spring-srv6-srh-compression-01} to perform load balancing. It first dynamically divides the satellites in between a lightly loaded zone, where there is little traffic, and a heavily loaded zone, where there is more traffic. For instance, densely populated area would map to heavily loaded zone. As there is little need for load balancing in the lightly loaded zone, a pre-balancing shortest path algorithm can be applied. In the heavy zone, a congestion index is used to find a minimum weight path so as to spread the traffic. Both zones use segment routing (SR) to implement the path selection. 

\cite{liu2015low} offers a low-complexity routing algorithm (LCRA) based on load balancing. It is a one-step distributed computation based on the location of the current node and that of the destination. Further, each node shares with its neighbors its congestion level, so that packets may be directed according to the congestion level of the links. 

\cite{dai2021distributed} performs the traffic allocation based upon the network topology (which density varies with the latitude) as well as the traffic characteristics. Indeed, it classifies the traffic into three categories: latency sensitive, throughput sensitive, and ordinary (best effort) traffic. Each traffic is routed accordingly so as to minimize congestion while satisfying the implied QoS requirements. 

\textbf{What next?} Absent data from satellite network operators, it is difficult to establish the practical need for load balancing mechanisms. It is obvious that traffic is not distributed uniformly, and that some parts of the network will have a higher utilization than others. It is also obvious that to provide better coverage, most networks will offer satellite-network path diversity between the ground stations: each ground station sees several satellites at all time, and each satellite may offer a distinct path if they are in different orbit planes. But we would need to know the utilization of these links to assess whether shortest-path routing leads to congestion issues. In any case, the idea (expressed in~\cite{dai2021distributed}) to offer different paths based upon traffic requirements seems a practical answer to this problem. 

\subsection{SDN-based Solutions}
\label{sec:SDN}

SDN~\cite{McKeown2008OpenFlow} has been proposed in 2007 and been successfully deployed in many data centers. Satellite networks are another type of networks where a logically centralized control can be deployed. As in the DC, the satellite network has well defined boundaries where such centralized control can be enforced. 

Further, satellite networks beg for the control to be offloaded into an independent control plane with more (cheaper) processing power (that is, on the ground) that can pre-compute routing and topology, and set simple forwarding rules in the satellites. 

\cite{yang2012SDS} is one of the earlier mentions of Software-Defined Satellite for Space Information Systems, but this brief note refers to software-defined radios. 

\cite{tang2014software} presents the challenges of using SDN for satellite networks, and designs an architecture. They leverage GEO satellites to collect information about the LEO layer and forward to the control (in the Network Operation Control Center (NOCC) on the ground. At the same time, \cite{bao2014opensan} made a similar proposal with OpenSAN, an SDN-based satellite network architecture which also involves GEO satellites to facilitate the control management. SERVICE~\cite{li2018service} follows a similar framework, with the addition of some NFVs and of the presentation of heuristics to achieve low latency or high bandwidth routing. 

As in the SDN-based architecture above, \cite{shi2019crossdomain} defines a framework with the main difference of integrating multiple layers into a Multi-Layer Satellite-Terrestrial Integrated Network, MSTIN.

\cite{ferrus2016sdn} also discusses SDN and NVF in the context of satellite virtual network operators to support multi-tenancy of satellite networks, with the goal of offering Satellite Networks as a Service, satellite-terrestrial hybrid services or cellular back-haul services.

\cite{papa2018dynamic} takes a different steps and does not assume the controller is on the ground, or in a GEO layers. Instead, it assumes some of the satellites are controller, and indeed, that the controllers can be instantiated within the satellite network based upon the demand. They formulate an optimization problem to provide the location of these controllers within the satellite network. 

\cite{zhu2017software} sets forth a proposal for a "master" controller and several "slave" controllers distributed around the earth (the terminology is from the paper; we would prefer more neutral terms). They propose a software defined routing algorithm (SDRA) that leverages the congestion in the satellite (as observed in the distributed controllers) in real time, to bypass the congested inter-satellite links when computing new paths. 

\cite{tang2020dynamically} formulates a SDN-based integration of ground and satellite networks, and defines an optimization problem to perform resource allocation and routing. This problem is NP-hard, and a network- and load-aware transmission protocol (denoted DEEPER) is proposed. 

Integrated Satellite-Terrestrial Networks (ISTNs) are the focus of~\cite{guo2019sdn}. This paper proposes a unified control architecture via SDN for both the terrestrial and the satellite networks to be managed jointly. An ISTN controller oversees a satellite network and a terrestrial controller. As an illustration, the paper shows how using some learning mechanism (ant colony optimization) can be leveraged to optimize the end-to-end path using fragment routing.  

\cite{wang2019adaptive} uses SDN to define a 3-layer satellite communication network model and implement an adaptive routing algorithm (ARA).

Using GEO satellites to assist the control plane, as mentioned above, may create bottlenecks~\cite{xia2019ground}. Therefore~\cite{xia2019ground}  reuses the ground stations instead of dedicated GEO satellites to establish a more scalable control plane. This introduces other issues, mostly that the ground stations have a limited view of the satellites, and this is addressed in the paper. 

\cite{boero2018satellite} looks at research to integrate satellite networks with 5G, in particular leveraging SDN. It presents an architectural framework, as well as open research challenges.

\cite{kumar2022fybrrLink} leverages the SDN framework to control the routing paths. Their proposal
makes use of Bresenham's algorithm ~\cite{bresenham1965algorithm}. This algorithm is used to find the nodes on a graph that approximate a line (it is used to draw lines with pixels on a screen), and they use it to identify the satellites that best approximate the path between the source ground station and the destination ground station.

\cite{jiang2023software} surveys the latest developments in software-defined satellite networks.

\textbf{What next?} As mentioned earlier, routing in satellite networks can use a {\em snapshot} approach or a {\em dynamic} approach. An SDN controller can be viewed as a hybrid mechanism: while the forwarding rules are not dynamic per se, they can be updated dynamically by the SDN controller. 

It seems to us that, from a practical point of view, the controller should be located on the ground and make decision for routing segments over the satellite path viewed as a single logical link, based upon current topology, congestion information and prior history. 

\subsection{Integration of Machine Learning}
\label{sec:ML}

Machine learning has been used to manage existing networks (see~\cite{boutaba2018comprehensive} for a comprehensive survey), and it is quite natural to apply these techniques to satellite networks as well. In particular, the high predictability of the topology, the high number of nodes, and the high variability of the network seem to point as appropriate conditions to apply ML methods. For instance, we believe that training a model under conditions that end up being periodic would lead to good results, as long as the training set comprises at least a multiple of such periods (the appropriate period varies on the satellite network topology as the nodes circulate around the Earth, as well as the periodicity of the traffic; one is of the order of hours and the other of days or weeks).

ML computations may be offloaded to some (logically and potentially ground-based) centralized controller and then we refer the reader back to the previous Section~\ref{sec:SDN}.

\cite{wang2019twohops} defines a Two-Hops State-Aware Routing Strategy Based on Deep Reinforcement Learning (DRL-THSA). It considers two-hop neighborhood information for a satellite, where the link state (codified into three levels) for the links within that neighborhood are considered. This two-hop link state is then given as input to a Double-Deep Q Network (DDQN) to figure out the optimal next hop. The paper evaluates the model setting, the training process and the running process.

\cite{na2018distributed} uses a form of feed-forward neural network that allows a fast training sequence (denoted Extreme Learning Machine, ELM). ELM is faster than other Artificial Neural Network (ANN) and is used to compute a distributed routing (ELM-DR) strategy. It takes as input the traffic density to estimate a traffic prediction; the routing is then derived from the traffic at each satellite node. 

\cite{rajagopal2021optimal} is another application of ELM to satellite networks. I applies a multitask beetle antennae search (MBAS) algorithm (for information on BAS, see here~\cite{jiang2017BAS}), using traffic prediction in combination with Mobile Agents to make routing decision. 

Q-learning is a model-free reinforcement learning algorithm to discover the value of an action in a particular state, and it has been used in networking for many years~\cite{littman1993distributed}. It was of course applied to satellite networks as well~\cite{huang2022reinforcement} with QRLSN, a Q-learning-based dynamic distributed Routing scheme for LEO Satellite Networks. QRLSN formulates a multi-objective optimization problems, and used the reward-based learning method to find a solution.

\cite{xu2022spatial} also uses reinforcement learning to design a Fully Distributed dynamic Routing algorithm based on Multi-Agent deep Reinforcement Learning (FDR-MARL). 

\textbf{What next?} This area seems emerging, with applications of a whole range of ML techniques, with no clear cut solution that is obviously better than others. The satellite systems has some properties that seem well suited for learning. Network management is a prime application area for ML and satellite networks are no exception. 

However, typical network management exports the data models in a manner that is yet to be supported in satellite networks. If inter-operability of satellite networks is bound to happen, then the data models will need to be updated to support these networks, and then be used as input to ML processing.

\subsection{Disruption tolerant networking \& Reliability}
\label{sec:disruption}

Most satellite network routing algorithms, such as snapshot-based (recall FSA~\cite{chang1998FSA} from Section~\ref{sec:history}) or dynamic (say, LAOR~\cite{papapetrou2007distributed}), are not particularly reactive to link failures. However, satellite network links are not stable and may be disrupted. As for one common example, consider the solar panels of the satellite  hiding the antennas for some period of the orbit. 

\cite{ji2015destruction} proposed DODR, a destruction-resistant on-demand routing protocol. This is a adhoc protocols that leverages route replies (RREPs in AODV) in the path discovery process, as well as a local repair strategy, to identify and fix broken paths quickly without incurring heavy overhead.

\cite{jin2022disruption} proposes a disruption tolerant distributed routing algorithm (DTDR). It reacts to failures by only broadcasting the failed link information (as opposed to link status updates) to reduce the overhead. DTDR takes advantage of the relatively static (or periodic) mesh
topology of the satellite constellation. It then combines a  static routing complemented with dynamic algorithms to efficiently compute the routing tables. 

\cite{qi2020distributed} designs a distributed survivable routing algorithm for mega-constellations with inclined orbits. Inclined orbits have different topology properties from polar orbits, but their predictability allows to determine multiple primary and secondary paths towards each destination. To handle failure, \cite{qi2020distributed} adds a recovery mechanism with  limited scope flooding and pre-computed detour mechanisms.

\cite{lu2014survivable}, mentioned in Section~\ref{sec:multilayer}, proposes mechanisms to handle link and satellite failures in a multi-layer context. 

\textbf{What next?} Failure recovery is an important component of networking. We believe that proper routing design with fast convergence may be the best option going forward. 

\subsection{Geographic Routing}
\label{sec:geo}

In 2000, \cite{henderson2000distributed} considered geographic routing, namely a routing mechanism that selects the next hop based upon minimizing the remaining Euclidean distance towards the destination.  In regular conditions, geographic routing is close to the optimal path (the paper bounds the delay difference by 10ms); however, it breaks down at the  counter-rotating seams, the polar regions, and close to the destination of a packet.

\cite{kumar2022fybrrLink} attempts to draw a straight line path through the satellite constellation towards the destination. We discuss this paper more in Section~\ref{sec:SDN}.

\textbf{What next?} Geographic routing has been considered in many dynamic topologies (for a couple examples, see \cite{herzen2011PIE,Westphal2009Scalable} in the case of ad hoc networks). The benefit of geographic routing is that the routing is extremely simple, once the relative position of the current node, its neighbors and the destination is known: picking the neighbor's closest to the destination is a simple computation. However, in a satellite network, the destination (on the ground) is not static with respect to the moving constellation (in the sky). The topology also varies between the ascending and descending satellites. This is practical as long as the neighbors have a relative motion that allows to calculate their position quickly. 

\subsection{Low latency Networking}
\label{sec:lowlatency}

Here we consider the end-to-lend latency of a packet in the network, between leaving the source and arriving at the destination (or alternatively, the Round-Trip Time, RTT). 
\cite{bhattacherjee2018gearing} makes the case that satellite networks will provide lower latency than current networks, when connecting points that are far apart (such as, say, a Frankfort-DC link). They even consider a ad-hoc network of planes (using existing flight paths data) to show the latency gain of relaying communications off flying objects at varying altitudes. 

\cite{lai2022spaceRTC} takes a similar tack by estimating the latency gain of a satellite path versus the meandering fiber path from the client to the server. It then proposes SpaceRTC, a mechanism to build an overlay with multiple close-to-optimal paths over the LEO constellation. 

\cite{zhang2022enabling} achieves low latency by heuristically solving a  Low-latency Satellite-Ground Interconnecting (LSGI) problem. They integrate ground and space segments to minimize the maximum latency, while at the same time, keeping the routing stable. 

Low latency using directed percolation \cite{hu2020directed} attempts to realize the URLLC 5G communication, which stands for ultra-reliable, low-latency communications. It does so by leveraging path diversity, and forwarding each packets twice to neighbors closer to the destination (one at the same orbit, one at a different orbit); the properties of the topology ensure that the packets will be forwarded over a narrow path where the intermediate routers can perform deduplication if they receive the same packet twice (once from a neighbor in their orbit, one from another orbit, for instance). The OPRAH protocol~\cite{westphal2006oprah} which defines a similar narrow path between source and destination seems relevant here.

\cite{geng2020optimal} takes not only delay, but also delay variations into account. It is based on the Graphical Evaluation and Review Technique (GERT). It obtains the minimal path set based on a delay queueing variant of GERT (DQ-GERT) then the delay index is computed that includes the delay and the delay variation on the path. The authors claim this is the optimal delay path. 

\textbf{What next?} Low-latency is one important driver for satellite networks. There is a huge premium in delivering some traffic faster. It seems that it is an intrinsic property of a well defined satellite network to be faster than a fiber or an oceanic cable. It also seems common sense that satellite will not carry huge buffers to introduce some congestion delays. The space for optimization, in addition to some basic QoS mechanisms, seems limited. 

\subsection{Multi-layer Satellite Networking}
\label{sec:multilayer}

\cite{xiaogang2016survey} surveys some of the proposals for routing across different satellite layers, between LEO, MEO and potentially GEO networks.

One of the earliest proposal was MLSR~\cite{akyildiz2002MLSR}, which stands appropriately for multi-layered satellite routing. MLSR considers LEO, MEO and GEO links and collects delays to create efficient routing tables. 

At about the same time, \cite{chen2002satellite} proposed another multi-layer approach, where LEO satellites were grouped together by the footprint of which MEO satellites they belong to. This creates a hierarchical partition where the MEO satellites are able to act as controllers to make routing decisions for the LEO satellites underneath, in what the paper calls the Satellite Grouping and Routing Protocol (SGRP).

\cite{lu2014survivable} expands the previous approach into a Survivable Routing Protocol (SRP) to include survivability and fast re-routing in case of satellite failure, at either the LEO and MEO layers. 

\textbf{What Next?} This area of investigation is important to integrate multiple layers, but seems to have slowed down. Current LEO satellite deployments are stand-alone, and not integrated with MEO or GEO (to the best of our knowledge). 

\subsection{Miscellaneous}
\label{sec:misc}

\textbf{Applications: } \cite{lai2021orbitcast} uses the LEO constellation to build a low-latency delivery network for Earth Observation (EO) application. This allows to download data from a distribution network in space without the low-bandwidth constraints of geo-stationary.

\cite{celikbilek2022survey} surveys the application of satellite networking to future autonomous transportation. Satellite networks offer a reachability that is well suited to transportation, but the gap between the mobility of the autonomous vehicle and the satellite has to be accounted for. 

\textbf{Energy Efficiency: } Energy-efficiency seems to be a natural consequence of satellite networks that is achieved for free: since satellite cannot be plugged or replenished with consumed fuel, they have to harvest the energy to function, mostly from solar panels. (Note: this does not account for the large amount of embedded energy spent into launching the network into space). 

However, \cite{yang2016energy} observes that energy efficiency is still a goal in these networks: since the number of charges of a battery is upper-bounded, the life span of a satellite is expanded by minimizing the number of charges/discharges. They therefore define an energy-efficient satellite routing problem, that is NP-hard to solve, and propose three heuristics to find approximation solutions: one baseline, one enhanced with a sleep cycle to turn off unused links, and one that takes into account QoS and link utilization. The last algorithm improves lifetime of a satellite battery by 40\% with little loss in performance. 

\textbf{Multipath:} the use of network coding and path diversity is known to improve the network capacity, especially in unreliable environment. This is adapted to the satellite network concept in~\cite{tang2019multipath}.

\textbf{Mobile Edge Computing: }
Mobile Edge Compute (MEC) has been deployed to provide computing and caching at the edge of cellular networks~\cite{MEC2015}. It therefore makes sense to extend this edge to the satellite networks, and~\cite{qiu2022mobile} surveys the relevant frameworks, concepts and challenges. As an example, \cite{hao2023joint} deploys mobile edge computing (MEC) servers in LEO satellites and jointly optimizes the computation offloading, the radio resource allocation and the caching
placement in LEO satellites.

\textit{ICN:} While satellites are typically assumed to have limited capability, some researchers have assumed that they will be able to cache content and support Information-Centric Networking mechanisms. For instance, ~\cite{yang2021towards} extends network coding in ICN~\cite{Montpetit2011Network,rfc9273} to satellite networks, and uses fountain codes and caching to reduce transmissions in an unreliable network. \cite{xu2022hybrid} categorizes satellites as core vs edge for content caching and retrieval purpose. 

\cite{wang2022caching} proposes to use caching in satellite networks, and to rank content and caches according to a spreading influence metric. For satellite, this means that satellites with high degree of connectivity are more likely to cache content, due to higher connectivity, and the higher connectivity of their neighbors (related in the paper to information entropy). In ICN networks, \cite{khan2020reversing} argues that caching should be at nodes with lower centrality (namely, lower node degree) so it is interesting to argue that the opposite is true in satellite ICN.

Similarly, \cite{zhu2023cache} uses centrality as a metric to select proper caching nodes, and as well seems to contradict the work on centrality in ICN from~\cite{khan2018nice,khan2018popularity,khan2020reversing}.

\cite{yan2023comparative} compares IP-based routing with ICN mechanisms, namely IP-based OSPF and Named-data Link State Routing protocol (NLSR). It reaches the conclusion that OSPF outperforms NLSR with less messaging overhead and faster convergence. This paper also studies the stability of the topology and the length of the snapshot.

\cite{diao2023low} adapts Named Data Networking for satellite networking by devising a fragmentation/reassembly mechanism: Direct Forwarding and Reuse of Fragments (DFRF). Fragment themselves can be cached, thereby reducing transmission delay.

\subsection{Simulation platforms \& Modeling}
\label{sec:simulation}

SaVi~\cite{SaVi} is a reference satellite constellation visualization platform that is open sourced. It has been available since 2006 and is continuously updated. The latest version, as of this writing, is 1.5.1a from January 2022. 

\cite{lai2020starperf} a mega-constellation performance simulation platform that enables constellation manufacturers and content providers to estimate and understand
the achievable performance under a variety of constellation
options. In particular, this platform captures the impact of the high mobility of the  satellites 
and estimates the network performance that is achievable from one area to another.

Hypathia~\cite{kassing2020hypatia} combines a python library with an ns-3 satellite simulator to generate LEO routing results. There are several NS-3 models for satellite networking evaluation, including~\cite{silva2017satns3} (upon which hypathia is built) or~\cite{schubert2022ns3leo}.

In addition to simulation, modeling is another tool to evaluate LEO satellite networks. An early paper, \cite{sun2004modiano} presents a simple model for routing in LEO constellations. It assumes that the network is a $N\times N$ mesh where each satellite is connected to four neighbors. This is a simple model, but allows for evaluation of different policies to resolve contention for transmission. Namely the paper looks at transmitting a packet at random; transmitting the packet first that has traveled the longest path; or transmitting packets on the shortest path. 

\cite{bhattacherkee2019network} first establish that the problem of inter-satellite topology design for a large LEO satellite constellations cannot be solved using integer programming, random graphs or ant-colony optimization. Their approach is to observe that the topology of such network offers a limited set of configuration for the local views of a single satellite, and this set of configuration is the same for all satellites. They call this potential local view at a given time a motif. The set of motif can be further pared down when the satellites are near their apex vs when they are near the equator. This facilitate setting up the topology, and \cite{bhattacherkee2019network} show significant improvement compared to a neighbor-grid connectivity, over both Starlink and Kuiper topologies. 

\cite{xiao2018LEO} proposes a Low Earth Orbit (LEO) satellite network capacity model and study the influences of the topology on the satellite network capacity. In particular, they show that using satellite going in opposite directions increases the number of potential links, and thereby the utilization of each link. 

\cite{wang2022stochastic} is the first study of satellite routing based on stochastic geometry. It considers a satellite position that is driven by a binomial point process, and derives some optimality results for routing on such node distribution. Note that most satellite constellations currently have a more structured organization. They also used stochastic geometry in other results~\cite{wang2022ultra,wang2023reliability}.

\cite{deng2021ultradense} computes the minimal number of satellites in a LEO constellations that is required to provide coverage and enough back-haul capacity for user terminals (UTs). Specifically, it takes the UT demands as input to generate the minimal number of satellites required to satisfy this demand. 

\cite{grislain2022rethinking} models the routing problem as an Unsplittable MultiCommodity Flow (UMCF) problem, with less congestion than the short-path algorithm that are commonly used in satellite networks. 

\cite{auddino2022nearest} focuses on selecting the first and last satellites on the path (and do not modify the routing once within the constellation). Due to properties of the satellite constellation topology (namely that a satellite will not be connected to all nearby satellites, but only some pre-defined neighbors in the next orbits), selecting the right starting and final points allows to select shorter, or more efficient paths, than connecting to the nearest satellite from the ground station. 

\section{Satellite Networking: Standardization Efforts}
\label{sec:standards}

\subsection{3GPP}
\label{sec:3GPPPsatellite}

3GPP is a major SDO (Standard Development Organization) that contributes to Satellite Networking from several angles, including Physical signaling, Radio spectrum, Radio Network, Use Case, Deployment, and System Architecture.

Initially, satellite network was treated as an extension to the terrestrial network and was considered only to provide service to areas where the regular terrestrial network would not be available. The satellite network should also provide services that are more efficient than terrestrial network, such as broadcasting services, delay-tolerant services, etc. After 5G however, the importance of satellites has grown in the 3GPP community. 5G has proposed to use Non-Terrestrial Network (NTN) to represent all networks that involve non-terrestrial flying objects, such as satellite network, high altitude platform systems (HAPS), and air-to-ground networks. Of all those networks, satellite networks are the major case, and others are special cases of satellite networks. 

There are two scenarios being discussed in 3GPP to integrate a satellite network with 5G, as illustrated in Fig. ~\ref{fig:satellite-network-5g}: LEO satellite network as 5G Access Network, and LEO satellite network as 5G Backhaul. 

\startfigure
\includegraphics[width=\columnwidth]{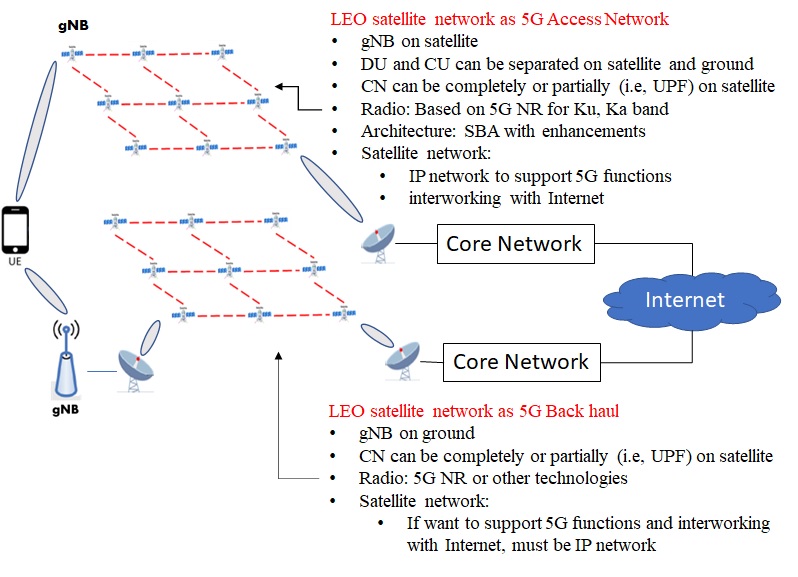}
\caption{Satellite network integrated with 5G}
\label{fig:satellite-network-5g} 
\end{figure}

Since Rel-15, 3GPP has proposed different Study items or Working Items in different TSG (Technical Specification Group): RAN (Radio Access Network), SA (Service \& System Aspects), CT (Core Network \& Terminals). The following TRs (Technical Report) were published: 

\begin{enumerate}
\item TR 38.811~\cite{TR.38.811}: “Study on NR to support non-terrestrial networks”, Rel-15
\item TR 38.821~\cite{TR.38.821}: “Study on solutions for NR to support non-terrestrial network”, Rel-16
\item TR 36.763~\cite{TR.36.763}: “Study on NB-IoT/eMTC support for NTN”, Rel-17
\item TR 22.822~\cite{TR.22.822}: “Study on using satellite access in 5G”, Rel-16
\item TR 23.737~\cite{TR.23.737}: “Study on architecture aspects for using satellite access in 5G”, Rel-17
\item TR 28.808~\cite{TR.28.808}: “Management and orchestration aspects with integrated satellite components in a 5G network”, Rel-17
\item TR 22.926~\cite{TR.22.926}: “Guidelines for extra-territorial 5G systems”, Rel-18
\item TR 24.821~\cite{TR.24.821}: “CT aspects of 5G architecture for satellite networks”, Rel-17
\end{enumerate}

3GPP expects the satellite network to directly connect to mobile devices and terrestrial network with acceptable bandwidth. Obviously this is a visionary feature and needs a lot of research and engineering work.  

The current regular mobile device (cell phone) cannot provide enough power to directly connect to satellites at an altitude of a couple of hundreds kilometers with a satisfactory speed to access Internet.  The traditional satellite phone can only provide the data rate about 10k bps, that is far below the expectation to obtain the Internet service. As a comparison, the data rate for StarLink service (with the use of a terminal ground-station that has about an antenna area of 50x30 \(cm^2\) and consumes about 50-75 watts on average) can provide couple of hundred Mega bps for down-link and tens of Mega bps for up-link. 

Therefore, some research will focus on the physical layer: how to design radio receiver on mobile devices and satellite that is super sensitive to the weak signal, and design transmitter that can transmit stronger signal under limited power supply. This may need revolutionary innovation in components, antenna, and semiconductor, etc. This is orthogonal to the scope of this paper, which focuses on the network layer. We only briefly mention radio issues next.

Since Rel-15, 3GPP has started the study for the NTN with New Radio (NR) technologies developed for 5G. In TR 38.811, Different aspects of NR for the use of satellite were studied. This includes: 

\begin{enumerate}
\item	Channel modeling for satellites when considering different user environments and atmospheric conditions.
\item	Satellite-specific constraints associated with satellite networks: Propagation channel; Frequency plan and channel bandwidth; Link budget; Cell pattern generation; Propagation delay characteristics and impacts; Mobility of transmission equipment and terminals; Service continuity crossing 5G and NTN; Radio resource management;
\end{enumerate}

Meanwhile, 3GPP also proposed architectures for the integration of NTN with terrestrial network under the assumption that the mobile device can connect with satellites directly. TR 38.821 for Release 16 described a satellite based NG-RAN architectures. In this proposal, the 5G architecture is used directly and satellites are treated as a complete or partial replacement for base station (e.g. gNB). There are three types of satellite in the report: 

\begin{enumerate}
\item	Satellite with transparent payload;
\item	Satellite with regenerative payload (gNB on board, with and without ISL - see below);
\item	Satellite with regenerative payload (gNB-DU on board, gNB-CU on ground - see below).
\end{enumerate}

The first type of satellite (see Figure~\ref{fig:transparent-mode}) represents the current work model for LEO satellite constellation such as StarLink: the satellite only does the signal relaying between ground stations. The only difference is that StarLink only uses its own ground station for terminal and Gateway, and uses its own proprietary technology instead of 5G NR for radio. For this type of architecture, there is no packet processing in satellite except the signal processing, such as Radio Frequency filtering, Frequency conversion and amplification. So, the base station functions are provided by devices on the ground behind the ground station. The corresponding control plane and data plane are shown in Figure~\ref{fig:CU-plane-transparent-mode}. 

\startfigure
\includegraphics[width=\columnwidth]{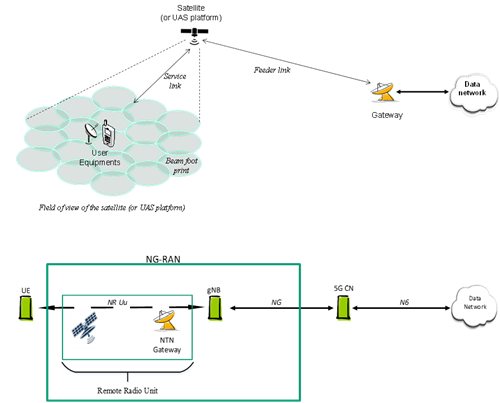}
\caption{Satellite with Transparent payload ~\cite{TR.38.821}}
\label{fig:transparent-mode} 
\end{figure}

\startfigure
\includegraphics[width=\columnwidth]{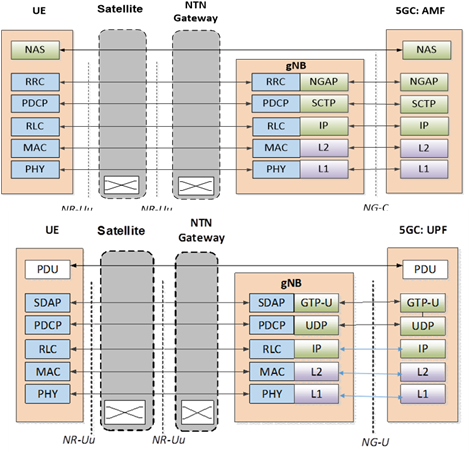}
\caption{Control plane and User plane for satellite with transparent payload ~\cite{TR.38.821}}
\label{fig:CU-plane-transparent-mode} 
\end{figure}

For the second type of satellite, in addition to the signaling processing function provided by transparent payload, the satellites also provide demodulation/decoding, switching and/or routing, coding/modulation. This is effectively equivalent to having all or part of the base station functions (e.g. gNB) on board the satellite (or UAS platform) as show in Figure~\ref{fig:regenerative-mode1)}. This is a general architecture for a satellite constellation integrated with 5G and Internet. Each satellite is functioning as a flying base station and the satellite constellation functions as back haul network, or core network. The satellite constellation connected by ISL will form an IP network and will be a carrier for the NG or Xn interfaces ~\cite{TS.23.501}. For this architecture, the AMF, UPF functions ~\cite{TS.23.501} are provided by devices on ground (see Figure~\ref{fig:CU-plane-regenerative-mode1}). 

\startfigure
\includegraphics[width=\columnwidth]{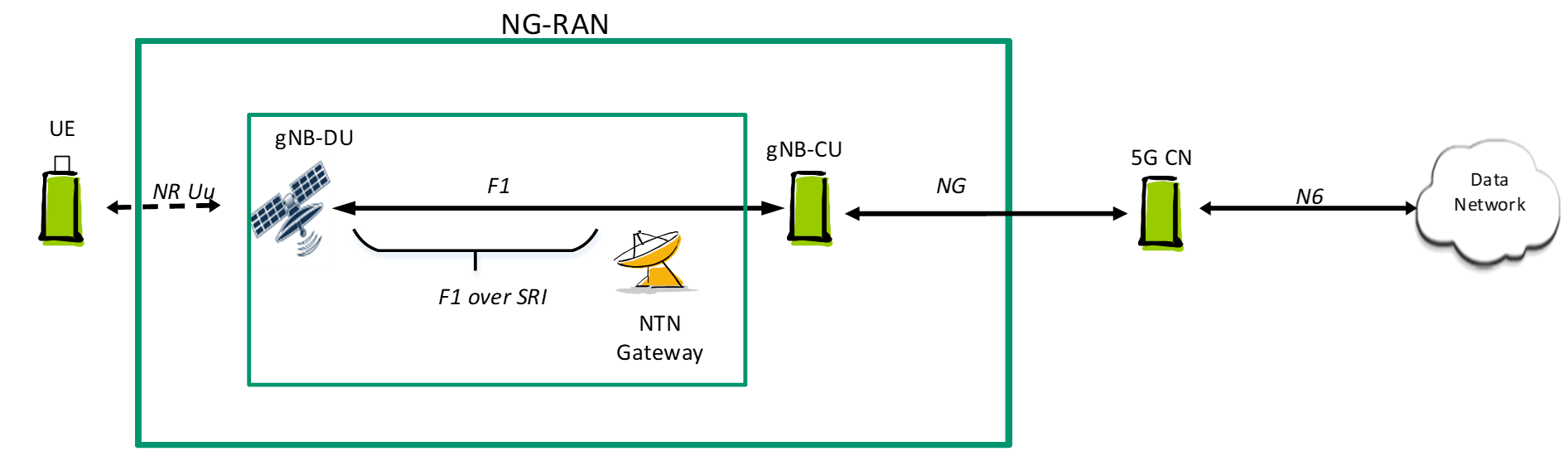}
\caption{Satellite with regenerative payload (gNB on board, with and without ISL) ~\cite{TR.38.821}}
\label{fig:regenerative-mode1)} 
\end{figure}

\startfigure
\includegraphics[width=\columnwidth]{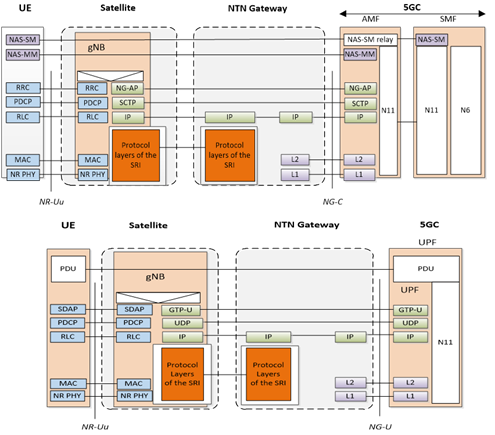}
\caption{Control plane and User plane for satellite with regenerative payload (gNB on board, with and without ISL) ~\cite{TR.38.821}}
\label{fig:CU-plane-regenerative-mode1} 
\end{figure}

The third type of satellite is similar to second type and is shown in Figure~\ref{fig:regenerative-mode2}, but each satellite will only provide part of the functions of base station. For this architecture, the control unit (gNB-CU) and data unit (gNB-DU) of the base station (gNB) are separated. The control unit of the gNB is provided by devices on ground; the satellite only does the data unit work. The user plane of gNB is also separated between satellite and device on ground (see Figure~\ref{fig:CU-plane-regenerative-mode2}) 

\startfigure
\includegraphics[width=\columnwidth]{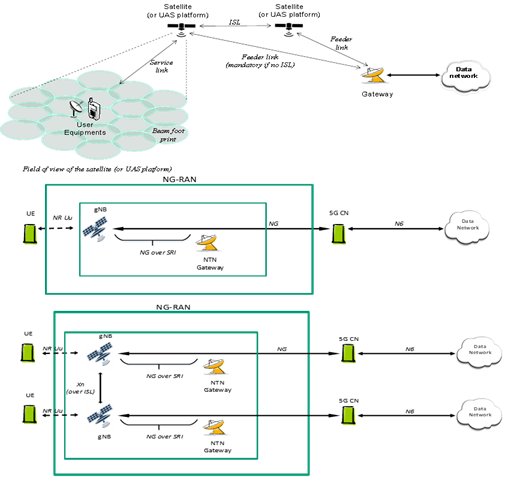}
\caption{Satellite with regenerative payload (gNB-DU on board, gNB-CU on ground) ~\cite{TR.38.821}}
\label{fig:regenerative-mode2} 
\end{figure}

\startfigure
\includegraphics[width=\columnwidth]{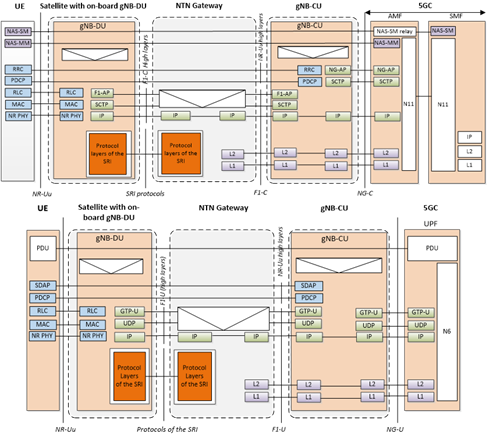}
\caption{Control plane and User plane for satellite with regenerative payload (gNB-DU on board, gNB-CU on ground) ~\cite{TR.38.821}}
\label{fig:CU-plane-regenerative-mode2} 
\end{figure}

Since the satellite network moves so fast (with the speed more than 7 km/s ~\cite{lhan-problems-requirements-satellite-net-02}), it impacts how NR is used. TR 38.821 also has done the detailed analysis and potential solutions. The analysis includes: 
\begin{enumerate}
\item Radio Layer 1 Issues: it analyzed about satellite parameters and UE characteristics for system level simulator calibration; Beam layout parameters for single satellite simulation, multiple satellite simulation. It also discussed about the Link level simulation, Link budget analysis; For physical control procedures, analysis for timing relationship, power control, beam management are done. Also the DL synchronization and Random access are discussed for uplink timing procedures.
\item Radio Protocol Issues, the report has analyzed the user plane enhancement for radio protocols like Media Access Control (MAC),  Radio Link Control (RLC), Packet Data Convergence Protocol (PDCP), Service Data Adaptation Protocol (SDAP). It also analyzed the enhancement for control plane in the areas: Idle model mobility, Connected model mobility, Paging, Radio link monitoring, Public land mobile network (PLMN) identities deployment and Ephemeral Data for NTN.
\item Architecture level and Interface Protocols issues: Tracking area management; Registration update and paging handling, Connected model mobility.
\end{enumerate}

For 3GPP's "Regenerative Payload" mode (used in the scenario where satellites networks provide 5G RAN services), the satellites must provide the functions of eNodeB (4G)~\cite{TS.136.101} or gNB (5G)~\cite{TS.123.501}; the devices on ground provide the packet gateway (PGW, in 4G terminology) and User Plane Functions (UPF, for 5G). IP connectivity within the satellite network is mandatory for the under-layer infrastructure. Both control plane and data plane are over the IP layer.

3GPP is still working on the scenario that satellite network in integrated with 5G. Two new projects "Study on 5G System with Satellite Backhaul" and "Study on satellite access, Phase 2" are under research (TR.23.700~\cite{TR.23.700}). Both specifications will be available in the next 3GPP release Rel 18. The project "Study on 5G System with Satellite Backhaul" focus on the scenario that uses the satellite network as backhaul network for 5G. The case is illustrated in Fig.~\ref{fig:satellite-network-5g} lower picture. The phone talks to a traditional 5G base station (gNB). The gNB is connected to a satellite network by a ground station. The satellite network is then connected to another ground station that is connected to Internet. This scenario does not need phone to communicate with satellite directly, thus is more realistic with the current technology and can be deployed quickly.

\subsection{IETF}
\label{sec:IETFsatellite}

As mentioned in Section~\ref{sec:IP}, the protocols of the current Internet are not appropriate for satellite networks. IP, TCP, BGP, OSPF and others have to be extended or modified to support running over satellite constellation networks. 
As a result, satellite networks have been considered in IETF for a long time as well. 

Originally, the popular satellite network like GEO only works as a radio or L2 links (bent pipe) between ground stations or satellite links are just like a physical links between two network nodes.Even the satellite link have special characteristics (longer delay and higher packet loss ratio), but it does not directly impact the higher layer networking protocol like routing and switching. Instead, only the higher layer transport protocol, i.e., TCP, needs to consider the impact of satellite link quality to TCP's behaviors.

The working group TCPSat~\cite{TCPSat} was created in 1997 to study the behavior of the TCP protocol over satellite links. It generated two documents, before closing in 2000: \cite{rfc2488} (Enhancing TCP Over Satellite Channels using Standard Mechanisms) and \cite{rfc2760} (Ongoing TCP Research Related to Satellites) which were published as Best Current Practice RFC in 1999 and Informational RFC in 2000, respectively. 

Recently, the LEO satellite, 3GPP's NTN project have attracted the attention of IETF. The new scenarios in NTN integration have raised the special requirements in routing area and have triggered some studies in IETF.

LISP~\cite{farinacci-lisp-satellite-network-01} is an IETF protocol and architecture that are currently considered to be used over satellite network systems.  The LISP overlay runs on the ground network, and uses the satellite network system as the underlay.

Time-variant routing was presented in IETF115 in 2022 as a BOF meeting which purpose is to establish a new WG (Working Group). The new TVR (Time-Variant Routing) WG has been established to define information and data models that address time-based, scheduled changes to a network. Time-based changes may include changes to links, adjacency, cost, and - in some cases - traffic volumes. Satellite network routing is definitely one of the main applications for a routing that evolves with time. Other applications would be to make the network greener based upon time-varying patterns in energy sources and energy costs. 

Further, as more and more satellite get into space, it is expected that they will need to inter-operate at some point. Therefore, defining a set of protocols for an Internet of satellites will be required in the near future. 

Recent work~\cite{lhan-problems-requirements-satellite-net-02,lhan-satellite-instructive-routing-00,lhan-satellite-semantic-addressing-01} motivates the need to specify some new routing protocols for large scale LEO satellite network by defining some requirements~\cite{lhan-problems-requirements-satellite-net-02} and specifying some proposals for instructive routing and semantic addressing for satellites. The routing solution ~\cite{lhan-satellite-instructive-routing-00} is actually using the source routing concept. By using semantic address of LEO satellite ~\cite{lhan-satellite-semantic-addressing-01}, the packet forwarding in satellite could be just a simple instruction like "forwarding the packet on specified direction until reaching a specified satellite". Compared with the traditional hop-by-hop IP packet forwarding (based on Longest Prefix Match lookup for the routing table), the new solution can be more adaptive to the very dynamic network topology while the packet overhead is less than the IPv6 Segment Routing (SRv6) ~\cite{rfc8402} that is also a source routing solution.
The new routing protocol is orthgonal to the path determination method, It could use modified OSPF as path determination protocol as in ~\cite{ospf-monitor} or any other methods.

In the IRTF, the potential for applying network coding to satellites was studied, including a taxonomy and a list of research challenges~\cite{rfc8975}.

The current satellite networks like StarLink are currently proprietary and are operated independently of the global Internet. 
However, limited domains~\cite{rfc8799} have been discussed to allow for the discovery of boundary nodes and to specify the inter-operation of limited domains with the wider Internet.

The  InterPlanetary Networking Special Interest Group (IPNSIG, https://ipnsig.org/) is a chapter of the Internet Society that works on networks in space. Out of their work came some protocols, and in particular, some Delay Tolerant Networking (DTN) proposals.  This led to the formation in IETF of a Standards Working Group. Formal adoption of DTN as a set of terrestrial Internet standards is in progress, including a few already published RFCs~\cite{rfc9171,rfc9172,rfc9173,rfc9174}.These are not LEO satellite related, but demonstrate that the IETF can standardized protocols designed for space. 

\section{Challenges and Future Research Directions}
\label{sec:challenges}

Satellite networks have obvious benefits in terms of coverage and global latency. However, there are some also downsides and challenges to solve. 

{\bf Challenge - Reliability: }For instance, a satellite orbit may provide useful coverage for only part of the time; the rest of the time is covering the poles or some empty areas such as oceans. For these areas, the satellite constellation may be too redundant, while for dense areas, it may be saturated. The issue of debris (for instance, as monitored by LeoLabs~\cite{leolabs}) in space has become an issue - as well as a business opportunity. LEO satellites are lightweight and move extremely fast, which makes any collision or impact challenging. For instance, 2,793 of the 3,055 satellites launched by Starlink were still in orbit as of early 2023, for a 10\% attrition rate. 
One challenge is to provide reliable networking in an environment that becomes more and more crowded and difficult. 

{\bf Challenge - Vertical Integration: }Another future challenge is to further integrate multiple layers: we discussed the interaction of LEO, MEO and GEO satellites already, but there are layers below as well. Some projects~\cite{singla2014Internet} have attempted to provide Internet at the Speed of Light - namely what LEO satellite networks attempt to provide, but doing so on the ground with microwave towers. This obviously cannot scale globally and combining some fast ground network over landmass with a satellite network over the oceans may be a viable solution. 

\cite{garcia2019direct} uses a network of UAVs to connect without going through a satellite network and staying within the atmosphere. They design a direct air-to-ground communications (DA2GC) system, and test it with a single UAV. This has the benefit of lower latency and higher throughput. The radio technology in such environment can be similar to 5G. 

\cite{cheng20226G} and \cite{Liu2018SpaceAirGroundIN} both survey the research that attempts to integrate the satellite network with not only the ground network but with an aerial network as well; integration with aerial networks may bring significant benefits, as they are somewhat complementary. There are few, if any, commercial deployments but it seems that a sparse global satellite network augmented with an aerial network over denser areas brings out the best of both worlds. 

The Alphabet project Loon~\cite{uyeda2022SDN} also used UAV (here, balloons). SDN was the technological solution to manage the network  infrastructure, and in particular the temporal evolution of the topology. It was shown to have applicability to space networks in~\cite{barritt2018loon}, even though the original project targeted low-connectivity areas in Africa. It was eventually spun out as a start-up, Aalyria (www.aalyria.com/) which offers SDN tools to manage LEO satellite networks as well. 

Integrating all these layers probably would require some dynamic framework to manage the network across all these layers, building upon SDN and machine learning. 

{\bf Challenge - Application support: }At the application layer, there are challenges to support new applications over LEO satellite networks. 

For instance, \cite{celikbilek2022survey} takes advantage of the ubiquitous coverage of SatNets to apply them to autonomous vehicles (as discussed above). Yet, such application requires stringent delay and reliability that is difficult to achieve through satellite networks. Providing reliable low latency communications over LEO networks is still an open challenge. 

An empirical study of the Starlink network~\cite{michel2022first} for instance finds a loss rate of 0.4\% in lightly loaded scenarios, and of 1.5~2\% in congested scenarios, as well as a delay that increases significantly from 50ms median RTT (light load) to roughly 100ms RTT (congested network). This study is from a single vantage point and for relatively short distance (from Belgium to Germany, for instance) but shows that high loss rate happens in current LEO deployments. 

{\bf Challenge - Interoperability: }We presented the current status of standardization efforts for satellite networking. There has been some work on giving virtual network operators hooks into the satellite network by using virtualization~\cite{abdellatif2016virtualization}, which is a step into opening satellite networks.

However, most satellites are currently incompatible between networks operated by different companies. Each companies manage their networks independently and there is little incentive for a large constellation operator to share that infrastructure with a smaller competitor. 

Further, current satellites have a limited number of ISLs. This is due to the fact that ISLs are complex: the distance between two satellites may reach 2,000 miles. Such links can be pointed at other satellites traveling in the same or neighboring orbital planes, since the relative motion is limited (similar speed, similar travel direction). This reduces the need to inter-operate, as one single operator may use up these few links to set up their own topology without having the need, nor the resource, to connect to other operators. 

However, this may be unsustainable. Increasing light pollution, produced by low Earth orbit (LEO) satellites, is becoming an issue for astronomers. Space debris is another issue where coordination is necessary. 

As the number of links on a satellite grows, this portends towards greater interoperability and the need towards global standards to maximize the utilization of the constellations up in space. 

\section{Conclusions}
\label{sec:conclusion}

We have presented a brief overview of the current state of Low Earth Orbit satellite networking. After a contextual background section, we looked at the latest research in this domain, with focus on load balancing, SDN, Machine Learning, reliability, modeling and others. 

We then focused on the state of the standardization, as more and more LEO networks get deployed and will need to inter-operate: with ground stations;  in between satellites in different networks; between terminals and networks; between other networks at different altitude, such as MEO, GEO, or on the ground or in the stratosphere (i.e. below 50km of altitude). 

We finally discussed some of the challenges that future research should tackle, with an emphasis on network reliability, cross-layer integration, application support and interoperability. 

\bibliographystyle{ACM-Reference-Format}
\bibliography{reference.bib}


\end{document}